**Conducting electrons in amorphous Si nanostructures: coherent interference and metal-insulator transitions mediated by local structure**

Yang Lu and I-Wei Chen

Department of Materials Science and Engineering

University of Pennsylvania, Philadelphia, PA 19104-6272, USA

**Without a periodic reference framework, local structures in noncrystalline solids are difficult to specify, but they still exert an enormous influence on materials properties. For example, thermomechanical responses of organic and inorganic glasses sensitively depend on the distribution of free volume or soft spots[1,2]. Meanwhile, strong electron localization[3] that endows unparalleled electrical breakdown strengths to amorphous insulators is easily compromised by local defects that promote inelastic tunneling over a variable range[4,5]. Here we report how metallic conduction can overcome strong localization in amorphous insulators of small dimensions, and how local structures can manifest their spectacular influence on such conduction. In amorphous Si, nanoscale electrons are so coherent that they exhibit robust quantum interferences reminiscent of the mesoscopic phenomena seen in weakly localized metal crystals[6]. Yet ultrasoft Si bonds emerge as the key local structures whose extraordinarily strong electron-phonon interaction coerces itinerant electrons into moving slowly at low temperature, even becoming trapped at all temperature when Si-O/N sites are provided. The local structures can be manipulated by a voltage or pressure to regulate charge storage, charge flow and metal-insulator transition. Also made**

**of Ge and oxides and nitrides, nanostructured amorphous conductors could offer opportunities for new applications.**

Electrical conductivity mostly depends on atomic bonding as evident from comparing conducting metallic glasses and insulating covalent or ionic glasses. But structure of amorphous networks does have an important effect. For example, whereas Si and Ge crystals can be substitutionally doped by electron donors or acceptors to induce an insulator to metal transition[7-10], similar doping of amorphous Si even at the heaviest level cannot ameliorate network's random potential. Therefore, despite a $10^{10}\times$ enhancement of room temperature conductivity[11], its value being limited by electron hopping reaches only 0.02 $(\Omega cm)^{-1}$, which is well below the minimum conductivity seen in metals. Other covalent amorphous networks such as chalcogenide $As_2Se_3$ also fall into strong localization at low temperature[4] with their electron wave functions decaying exponentially not able to reach much beyond a localization length $\zeta$. The highly polarizable amorphous chalcogenides additionally have one peculiar feature of particular relevance to this study: Their same-state paired electrons are so strongly coupled to the atomic motion that they experience effective attraction with each other[12,13]. With this background, it came as quite a surprise to see metallic behavior in a large variety of ~10 nm amorphous thin films made of amorphous $Si_3N_4$ co-sputtered with a few percent of atomically dispersed metal atoms (Cr, Cu, Ta, Pt and Al.)[14,15] Also surprising is to see them exhibit characteristics of strong electron-phonon interaction, similar to that in chalcogenides, even though they have rather low polarizability. Other similarly prepared, Pt-doped, amorphous films of $SiO_2$, $MgO$, $Al_2O_3$, $HfO_2$, $Ta_2O_5$, $Y_2O_3$ and AlN confirmed the same observations[15], suggesting a general possibility for electron to overcome strong localization while experiencing elevated electron-phonon coupling in amorphous nanostructures. Could it be related to the local structures that somehow become more prominent in a nanoscale setting? Can the same local structures explain why almost all inorganic glasses—with the exception of metallic glasses—are insulators?

Since Si and Ge are arguably the best understood amorphous materials, we prepared sputtered amorphous films to answer these questions. We asymmetrically sandwiched the films between a continuous Mo bottom electrode and a patterned Pt top electrode to define many Pt/Si/Mo cells of various areas (inset in **Fig. 1a**; Pt has a higher work function than Mo.) We verified the amorphous structure by diffraction (**Extended Data Fig. 1a**), and measured the two-point resistance across the cell in the small-signal, Ohmic regime (**Extended Data Fig. 2a-b**, also see **Method**.) The resistance in **Fig. 1a** is strongly dependent on film thickness. Films 25 nm and thicker behave like a typical insulator, becoming vastly more resistive when cooled to low temperature ($T$) before the resistance finally saturates, apparently due to tunneling. Films thinner than 25 nm behave like a metal with nearly temperature-independent resistance. (Thus, the standard four-point measurement with a long distance between probes cannot reveal metallicity in our samples.) Very similar thickness-mediated conductivity transitions were also seen in other amorphous films, such as n-doped Si and p-doped Si (**Extended Data Fig. 2c-d**), as well as Si/Ge films with additionally sputtering-incorporated N or O (e.g., n-Si+N and Ge+O films in **Extended Data Fig. 2e-f**, the N/O content determined by x-ray photoelectron spectroscopy in **Extended Data Fig. 1b-e**.) Moreover, as will be detailed later, N/O-doped amorphous Si and Ge films can also be electrically or mechanically stimulated to undergo insulator-to-metal transitions—to multiple metallic states of different resistance values—at both 2 K and 300K. Therefore, this amorphous nanomaterial family offers a very rich landscape for our study.

The extrapolated 0K resistance provides strong evidence that the thickness-mediated conductivity transition is caused by localization. In **Fig. 1b**, $R_{0K}$ rises exponentially with thickness $\delta$ from which we use $R_{0K} \sim \exp(\delta/\zeta)$ to estimate $\zeta = 4.9$ nm. This is $18\times$ the Si-Si bond length of 0.27 nm, thus a surprisingly long distance. In the insulator regime, the flat tunneling resistance crossovers to Mott variable-range hopping above certain $T_{\text{crossover}}$, following his $\exp((T_o/T)^{1/4})$ law[16] with $T_o \sim \zeta^3$ as shown in **Fig. 1c** (also in **Extended Data Fig. 2c,** inset). The law explains the high temperature data in **Fig. 1a** in thicker films, where hopping occurs between localized states separated by various longer-than-$\zeta$ distances that overall scale with $T^{-1/4}$. When the hopping distance reaches the film thickness, the crossover happens. Thus, $\delta \sim T_{\text{crossover}}^{-1/4}$, which is also confirmed by the inset of **Fig. 1c**. These results are in agreement with the past studies of bulk amorphous Si and Ge in the strong localization regime[16].

We now describe weak localization in the metallic regime. In a nominally metallic disordered electron system, there are various electron interference phenomena giving evidence for weak localization as a precursor to strong localization[17,18]. Excluding spin-orbit interactions and magnetic scattering as unimportant in our materials, we may attribute these phenomena to *enhanced* electron-electron interaction (EEI) and *singular* backscattering by impurities/defects. They arise because electrons in a random network cannot travel ballistically, instead they must random-walk and diffuse. Therefore, after one encounter, two electrons may meander around in the "maze" and stay close to each other for some time before they finally diffuse away, and such "stickiness" enhances interaction. At 0K and the zero-energy limit, an incident diffusive electron can be Bragg-scattered off the Friedel oscillations of other electrons, giving rise to singular backscattering[19]. But singular backscattering also arises from diffusion of non-interacting electrons because they have a certain probability to diffuse back for a second encounter, which amounts to singular backscattering—if they have not lost coherency already. Previously, weak localization phenomena were mostly studied in metal samples with a measurement distance that is either macroscopic or mesoscopic[6]. But in our metallic samples that have a transport length of only ~10 nm~$\zeta$ (**Fig. 1c**), we have found two signature features of EEI-dominated weak localization phenomena: resistance ($R$) minimum in **Fig. 2a-b** and positive magnetoresistance in **Fig. 2c-d**. Furthermore, they exhibit thickness-dependent resistance saturation at low temperature in **Fig. 2e**, which is uncommon for mesoscopic phenomena because the typical sample sizes used in the past were too big.

The EEI-dominated weak localization theory predicts a quantum correction to three-dimensional (3D) conductivity[18],

$$\Delta\sigma(T,B)=\sigma(T,B)-\sigma_o=\alpha(4/3-3\tilde{F}_\sigma/2)\sqrt{T}-0.77\alpha\tilde{F}_\sigma\sqrt{T}g_3(\tilde{h}) \qquad (1)$$

Here, $\sigma_o=\sigma(0,0)$, $B$ is magnetic field that causes a normalized Zeeman splitting $\tilde{h}=g\mu_B B/k_B T$ (we shall set $g=2$), $g_3$ is an analytic function of $\tilde{h}$, and the two material parameters are (i) $\alpha=(e^2/\hbar)(1.3/4\pi^2)(k_B/2\hbar D)^{1/2}$, expressed in terms of the diffusivity $D$ that characterizes random walk, and (ii) $\tilde{F}_\sigma$, a Fermi liquid constant that characterizes the strength and sign of electron-electron interaction. These two parameters can be determined from the temperature and field dependence of $\Delta\sigma/\sigma_o$, which equals the normalized resistance correction $-\Delta R/R_0$ from $R_{0K}$, in **Fig. 2b-d**. The resistance saturation in **Fig. 2e**, which signals the diffusion distance is approaching the film thickness, can next be used to determine $D$, hence $\alpha$. Below we outline the analysis for n-Si:0.45N nanofilms of various thickness (8-16 nm)—and for 11 nm films, of various metallic states (28-393 Ω).

The main panels of **Fig. 2** display the temperature/field responses of $\Delta R/R_0$. Here, each overlapping data plot contains multiple data sets, either from several metallic states (**Fig. 2a-d** and their insets) or from several thicknesses (**Fig. 2e** and its inset). Indeed, it could have included many more sets of magnetoresistance data in **Extended Data Fig. 3** because they all fall onto the same $\Delta R/R_0$ plot. Therefore, $\Delta\sigma/\sigma_o$ of this material is state and thickness independent. As described in **Method** and Ref. 20, because of this independence, $\Delta R/R_0=-\Delta\sigma/\sigma_o$ can be rigorously extracted from a set of two-point resistances, measured for films of different thickness or different metallic states, from which the contributions of the electrodes, the loads, and the interfaces can be removed. Moreover, in the main panels and the insets of **Fig. 2b-d**, we have shown the theoretical predictions as solid curves using $\tilde{F}_\sigma=0.75$ and $\alpha/\sigma_o=5.76\times10^{-3}$ to illustrate their very good agreement with the data. They include: (i) from the first term of Eq. (1), the predicted $T^{1/2}$ dependence at low temperature for $\Delta R/R_0=-\Delta\sigma/\sigma_o$ (**Fig. 2b**), (ii) from the second term with $g_3(\tilde{h})$, the predicted $B^{1/2}$ dependence at high field (**Fig. 2c-d**) and the $B^2$ dependence at low field (inset of **Fig. 2d**) for magnetoresistance. Lastly, magnetoresistance is independent of the field orientation (**Extended Data Fig. 3i-k**), which attests the 3D nature of electron conduction in these films. Altogether, the theoretical predictions with the same values of ($\tilde{F}_\sigma$, $\alpha/\sigma_o$) were found to agree with about 50 sets of data plots in **Fig. 2b-d** and **Extended Data Fig. 3**, collected over a range of temperature (18 mK to 2K), field (−18-18 T), field orientation (0-180°), film thickness (8-16 nm) and metallic states (20 different resistances in the 11 nm film.) A similar analysis for several other film compositions also established the state/thickness independence and good agreement with theoretical predictions (solid curves in **Extended Data Fig. 4**) when using their respective ($\tilde{F}_\sigma$, $\alpha/\sigma_o$) values in **Extended Data Fig. 4a**.

The resistance minimum seen in the less resistive metallic films of 11 nm n-Si+N (**Fig. 2a**) and others (**Extended Data Fig. 4d-e**) can now be understood. At higher temperature, the resistance increases because incoherent (Drude) electrons increasingly suffer from phonon scattering. At lower temperature, the resistance again increases because coherent electrons increasingly suffer from interference. The latter's $T^{1/2}$ dependence displayed in **Fig. 2b** and **Extended Data Fig. 4a** comes from the temperature dependence of *coherent* diffusion distance, $L_T=(D\tau)^{1/2}=(\hbar D/k_B T)^{1/2}$, which is set by electron's dephasing time, being $\hbar/k_B T$ at low temperature. This further confirms 3D diffusion, because 1D and 2D diffusion would have given entirely different scaling laws (**Extended Data Fig. 4f**)[18]. Such $T^{1/2}$ dependence eventually breaks down when $L_T$ approaches the film thickness $\delta$, which results in resistance saturation. For n-Si+N, we illustrate this in **Fig. 2e** by displaying the data for different thickness against normalized coherent diffusion distance, $L_T/\delta \sim 1/(\delta T^{1/2})$. Taking the bend-over at $\delta T^{1/2}=4.8$ (nmK$^{1/2}$), we obtain $D=3.00\times10^{-6}$ m$^2$/s in **Fig. 2e** and essentially identical results for other metallic states of the same thickness (11 nm) shown as empty symbols in **Fig. 2e** and in **Extended Data Fig. 4i**. Corresponding data (**Extended Data Fig. 4g-h**) for n-Si and Si films also give similar $D$ values (**Extended Data Table 1**). Importantly, they are fall well below the lower limit $D=10^{-4}$ m$^2$/s of past studies. Incorrectly as we shall see, one might naively take this as caused by the much shorter mean free path $l$ for random walk in amorphous Si, for $l$ is set by the much longer impurity spacing in metals in the past studies.

Interestingly, despite the saturation of zero-field resistance below the saturation temperature $T_s \sim 0.2$K, the positive magnetoresistance in **Fig. 2d** continues to follow the $(T,B)$ scaling and $g_3(\tilde{h})$ but the curve can no longer be fit if one uses $g_3(\tilde{h})$'s $T^{1/2}$ prefactor in Eq. (1). This provides another confirmation of EEI, because if there were no EEI, then capping the diffusion distance ought to also cap singular backscattering, hence saturating the magnetorestance. (Singular backscattering without EEI contributes less to quantum resistance correction than EEI, because self-crossing paths, which are very common in 1D and 2D, are relatively scarce in 3D.[19]) We reason that the $T^{1/2}$ prefactor of the $g_3$ term in Eq. (1) should be replaced by $T_s^{1/2}$ once the coherent diffusion distance is capped by $\delta$. So we divide the data in **Fig. 2d** by a suitably chosen $T^{*1/2}$ to bring the quotient to conformity with the $g_3(g\mu_B B/k_B T)$ scaling of Eq (1), and thus chosen $T^*$ is plotted in the lower inset of **Fig. 2d**. Indeed, it becomes flat below $T_s \sim 0.2K$, which provides an independent check for the value of $T_s$. Lastly, because electrons are always localized at 0K in 1D and 2D[21], crossover from 1D/2D weak localization to strong localization should set in at a finite temperature when the phase coherent length exceeds $\zeta$.[22] In our samples $\delta \sim \zeta$, so the $L_T \sim \zeta$ and $L_T \sim \delta$ conditions nearly coincide. Therefore, the fact that resistance saturation instead of divergence was found at $L_T \sim \delta$ again implies 3D conduction.

Because the quantum correction is usually small for 3D conductivity, $\sigma_o$ should not be far from the Drude conductivity, which for electrons of a density $n$ and a Fermi wave number $k_F$, is $\sigma_D = (e^2/\hbar)(n/k_F^2)(k_F l)$. For a spherical Fermi surface, $n = k_F^3/3\pi^2$. So if we let $k_F = \pi/l$, we obtain $\sigma_D = (e^2/\hbar)/3l = 811.4(\Omega cm)^{-1}/l(nm)$, which according to Mott is the minimum conductivity for metal, being 3,005 $(\Omega cm)^{-1}$ if $l=0.27$ nm. Our $D$ and $\alpha/\sigma_o$ in **Extended Data Table 1** give a comparable value for $\sigma_o \sim 2,000$ $(\Omega cm)^{-1}$, confirming the smallness of the correction; it also does not contradict setting $l$ at about the Si-Si distance. Using $D = l^2/3\tau = lv_F/3 = k_F l\pi\hbar/3m^*$, where $\tau$ is the time of flight for random walk and $v_F = \hbar k_F/m^* = l/\tau$ is the Fermi velocity of electrons with an effective mass $m^*$, we find $m^*$ must be at least 30 times the rest mass of electron, for wavelike electrons must have $k_F l > 1$. So the very small $D$ revealed by our experiment actually arises from a very large $m^*$ instead of a very small $l$. Obviously, the Fermi velocity is also very low, and **Extended Data Table 1** gives estimates of $v_F$, $m^*/m_e$, $k_F$, free electron density $n$, and other electron properties for several materials studied here assuming $l=0.27$ nm, and $\sigma_o=\sigma_D$ (See **Methods** for the details of calculation).

Heavy electrons are usually due to the Kondo effect[23], which exists in nanostructures[24], but is ruled out here because our samples have no magnetic impurity or heavy atom. Another possibility is the narrowing of band width due to poor orbital overlap, or the compression of the valence/conduction band edge by the insertion of localized states[25], but this effect is unlikely to be so large to account for the >30 fold increase of $m^*$. On the other hand, the Frohlich Hamiltonian that involves a very strong electron-phonon interaction $\phi_{ep}$ of a coupling strength $\lambda$ predicts localization of an electron with a "band" mass $m_b$ into a polaron with a mass $m^* \sim 0.0202\lambda^4 \, m_b$.[26] So if a very strong $\phi_{ep}$ exists at certain local structures, such as ultrasoft Si-Si bonds, they could act like impurities and a virtual bound state is induced on conducting electrons. Using the polaron prediction for estimate, our data would suggest $\lambda \sim 6.2$, which is extraordinary large compared to $\lambda \sim 3.7$ in highly polarizable CsI and SrTiO$_3$ crystals[26]. But as we shall see next, in mechanically triggered electronic transitions our samples do give direct evidence for a very large but localized $\phi_{ep}$. Before going to such observations, however, we should note that while $\sigma_o \sim 2,000$ ($\Omega$cm)$^{-1}$ is $\sim 10^{14}$ times the room-temperature conductivity of undoped amorphous Si and $\sim 10^4$ times that of the most heavily (P and B) doped amorphous Si[11], the nominal conductivity ($\sim 5 \times 10^{-5}$ ($\Omega$cm)$^{-1}$) of our films is much lower[27]. Therefore, the metallic phase with slow but robustly coherent electrons is nested in a very small fraction of our films, but it does form contiguous conducting pockets/pathways between electrodes of these thin films.

We now describe electrically and mechanically triggered transitions in O/N-doped Si/Ge films of an intermediate thickness, and argue that they are caused by localized $\phi_{ep}$ and a negative correlation energy ($U$) at ultrasoft spots. To set the stage for discussion, we present in **Fig. 3a** an example of voltage-triggered transition in an n-Si+N film, but very similar transitions occur in **Extended Data Fig. 5** in films of another 11 compositions, including their as-fabricated, untested films as well as films that were subject to various pretreatments as elaborated below. In all cases, a positive ~ 1 V applied to the Pt electrode can transition the film from a metallic state to the insulator state, which retains $10^5$-$10^8$ Ω after the voltage retreats to zero. Next, the insulator state can transition to a metallic state, or any number of metallic states of an intermediate resistance, when a negative ~ −1 V is applied to the Pt electrode, as shown in **Fig. 3a** and **Extended Data Fig. 5.** (We used these metallic states to collect the data for **Fig. 2**.) Lastly, the same ~ ±1 V can trigger transitions over a wide range of temperature (**Fig. 3a**). Their transition curves have similar metallic states but the insulator state is much more resistive at 2K than at 300K (also see **Extended Data Fig. 5l**.)

The connection to electron-phonon interaction is established by mechanically triggering the insulator-to-metal transition using two interchangeable means. One is a hydraulic pressure $P_\mathrm{H}$ lasting for ~1 minute (**Fig. 3b**), the other is a magnetic pressure $P_\mathrm{B}$ lasting for $10^{-13}$ s (**Fig. 3c**)[15]. (Two examples of macroscopic $P_\mathrm{B}$ are (a) in the wall of a high-field magnet and (b) in the aircraft launch pad on USS Gerald R. Ford[28].) Below we will give some details of these experiments, previously reported for Cr-doped amorphous $Si_3N_4$ thin films[15]. (Readers may elect to skip this part.)

*The $P_\mathrm{H}$ experiment*: The experiment was performed inside a pressure vessel and the pressure was exerted from outside the electrodes against the film (see schematic in **Fig. 3b**). As $P_\mathrm{H}$ increased from 2 MPa to 1 GPa, it resulted in a higher transition yield and a lower average resistance for Si+O in **Fig. 3b** and for another 3 compositions including n-Si+N in **Extended Data Fig. 6.** Regardless of $P_\mathrm{H}$ used, the metallic state thus rendered can transition back to the insulator state with a positive voltage as shown in **Extended Data Fig. 8,** which proceeds in the same way as in **Extended Data Fig. 5**. Moreover, the pressure-transitioned films have the same $\Delta R/R_0 = -\Delta\sigma/\sigma_o$ response as voltage-transitioned films, at all $(T, B)$. For example, in **Fig. 2e**, the curve of resistance saturation of a pressure-transitioned metallic state (yellow) overlaps with that of a voltage-transitioned metallic state of the same composition and thickness (blue); their 18 mK relative magnetoresistance curves in **Extended Data Fig. 3d** also overlap.

*The $P_B$ experiment*: The experiment was performed inside a linear accelerator (SLAC) that shot a single bunch of $10^9$ 20 GeV electrons at the film, just once. The bunch has a size of 20 μm×20 μm×20 μm and carries with it a pulse of circumferential magnetic field of $10^{-13}$ s, which is the time to travel 20 μm at ~the speed of light[29]. The field peaks at 65 T at the edge of the bunch ($r$=20 μm), falls off as $\sim 1/r$ with the radial distance $r$, and remotely delivers to a two-side-electroded film (located at $r$ away from the flight path) a $P_B$ that peaks at 1,680 MPa and decays with $\sim 1/r^2$. (There is also an induced pressure due to the induced current in the electrodes, which inversely depends on the (pulse width)$^2$.)[15] In the schematic of **Fig. 3c**, the pressure is actually a body force on the electrodes. But to keep the upper electrode in balance an interface uniaxial tension of the same magnitude as $P_B$ must be generated, which will also stretch the film as if there is a negative pressure. These forces in turn generate a biaxial tension in the top electrode, which can be ripped apart as in the earth-colored region in the center of the right panel of **Fig. 3c**. (Rupture can be avoided by using a thicker, hence stiffer, top electrode[15,30].) Like the case of $P_H$, a higher transition yield and a lower average resistance result when $P_B$ increases, as seen with decreasing $r$ in **Fig. 3c** for n-Si+N and in **Extended Data Fig. 7** for the other two compositions. (In **Fig. 3c**, the two cells at $r$=0 experienced no transition because $P_B$=0, by symmetry.) Again as in the case of $P_H$, the metallic state thus rendered can transition back to the insulator state with a positive voltage and same transition curves indistinguishable from those in **Fig. 3a** (See **Extended Data Fig. 8b-d**).

These mechanical experiments provided definitive evidence for electron-phonon interactions in real time: As much as a $10^5\times$ drop in resistance is triggered by a very modest (positive or negative) mechanical pressure lasting from 0.1 ps to 1 minute in films that are not in contact with any electrical voltage. Note that both voltage- and pressure-triggered transitions have identical outcome in subsequent transition curves and the $\Delta R/R_0 = -\Delta\sigma/\sigma_o$ response to ($T$, $B$). However, although voltage can trigger the transition in both directions, pressure can only trigger the insulator-to-metal transition. Importantly, for the transition to occur there must be Si-O/N-Si bridges—we have not seen resistance transition in films that have not incorporated some O or N. To explain these results, we will adopt the energy-configuration diagram (**Fig. 3d**) of Street and Mott for spin-paired bipolarons in amorphous chalcogenides[13], but we will specialize it to the vicinity of an ultrasoft Si-O/N-Si bridge in amorphous Si because, unlike chalcogenides, neither Si nor $SiO_2$ is highly polarizable.

We now apply the idea of a localized negative-$U$, which we previously proposed to explain resistance-switching in other amorphous materials—metal-doped oxides/nitrides[15,31], to electron attachment/detachment to a strained Si-Si bond next to a strained Si-O-Si bridge in **Fig. 3d**:

(i). The initial ground state (green) contains a strained Si-Si and Si-O-Si bond.
(ii). The activated metastable state prior to relaxation (grey) contains an extra electron burdened by a positive on-site Coulomb energy, i.e., Hubbard-$U$.

(iii). The relaxed stable state (blue) contains a Si-Si triple bond, having a longer bond length with two electrons in the bonding σ orbital and one electron in the anti-bonding σ* orbital, overall relaxed and stabilized by an effectively negative $U$.

(Other possibilities may also apply. For example, with the addition of one electron and relaxation, the strained Si-Si bond may break to form a Si dangling bond and a Si with one set of lon-pair electrons. Alternatively, the strained Si-O-Si bond may break to form a Si dangling bond and an O with one set of lone-pair electrons.) To turn the positive Hubbard-$U$ in (ii) into the negative $U$ in (iii) after bond severing, this Si-O neighborhood must be a "soft spot" where the newly formed bond(s) can easily lengthen or recoil from each other, and from the lone-pair electrons, thereby lowering the electron repulsion, without incurring much elastic energy. It also helps to have an initially highly strained Si-Si and Si-O bonds that arise from O incorporation. In **Fig. 3d**, this is represented by a large *local* $\phi_{ep}$ that overcompensates the positive $U$, and the amount overcompensated is the negative $U$ that stabilizes the trapped electron. Assuming the trapped electron can exert Coulomb blockade to itinerant electrons, we shall associate (i) to an unoccupied trapped site in the metallic state and (iii) to an occupied trapped site in the insulating state, which explains the metal-to-insulator transition triggered by a voltage that brings in an extra electron, which becomes trapped. The reverse transition can then be explained by forcing configuration (iii) to configuration (ii), either electrically or mechanically, so that the trapped electron must leave. In this picture, there is no pressure-triggered metal-to-insulator transition (**Extended Data Fig. 7c-d**) because there can be no pressure work on an undistorted configuration (i).

The picture is further consistent with the following observations. (a) While amorphous structure always contains some very soft local structures, their population is low and their distribution is statistical[2,32]. This explains why very low pressure suffices in our mechanical experiments and why the transition yield is statistical, increasing with pressure in **Fig. 3b-c**. (b) The very low overall electron density and the very small conducting fraction in amorphous Si make it possible for a stabilized trapped electron to block itinerant electrons by long-range electrostatic repulsion. As long as some unaffected, far-away conducting paths remain, conduction can continue in a self-similar way, which explains why different metallic states can share the same $\Delta R/R_0 = -\Delta\sigma/\sigma_o$. The insulator state finally obtains when trapped electrons have blocked all the contiguous conducting paths. (c) Conversion from configuration (iii) to configuration (ii) may take as short as one atomic vibration (~0.1 ps). This explains why our 0.1 ps $P_B$ experiment works. Transitions at a much lower nominal rate should be rate-insensitive. Indeed, we were able to voltage-trigger transitions from 2K to 300K, and with voltage pulses lasting from 1 ms to 1 ns, all at the same critical voltage ~ ±1 V. (d) While a voltage of ±1 V can cause transition, the negative $U$ in (iii) can also explain why without such voltage the insulator state is stable: We found undisturbed insulator states can remain for years. (e) Lastly, it is conceivable that strained Si-Si bonds at low-density spots in O/N-free Si can likewise have a strongly localized $\phi_{ep}$, and cause a virtual bound state on a conducting electron thereby increasing latter's effective mass. But lacking the O/N heterogeneities that create further distortion to help immobilize electrons, O/N-free Si cannot undergo stimuli-triggered metal-insulator transtion.

While the ultrasoft, thus highly polarizable Si-O vicinity has a strong effect on conductivity, its fraction is probably too small to lower the dielectric constant of our films. (Likewise, we expect a very low population of ultrasoft Si-Si bonds.) This should be obvious from the very low conducting fraction in our films, and it was confirmed by measuring the capacitance $C$ of the insulator films using AC impedance spanning over $10^2$-$10^7$ Hz, which gave similar result as reported in the literature. However, when the resistance is lowered in **Fig. 3e**, we found a broad correlation between decreasing $C$ and decreasing $R$, the latter tuned by either applying a constant positive voltage to the Pt electrode to lower the resistance without causing transition, or by applying a transitory negative voltage to transition to various metallic states in films that contain O or N. In all, when the resistance decreases from 800 MΩ to 500 Ω, the capacitance decreases by ~ 50%. The correlation is reversible, which is verified by taking advantage of the "volatile" resistance of a positively biased insulator film (**Fig. 3a**). As the resistance returns to the original resistance when the voltage is removed, the capacitance also returns to the original value. (A negative-bias-triggered transition is non-volatile and will keep the new resistance and the new capacitance.) The large capacitance drop can be explained if many insulating dielectric pockets are short-circuited around—they can no longer be charged when the electrodes are biased. Under a positive bias, we envision a large electric field may lower the random potential to facilitate conduction, thus lowering both $R$ and $C$. But if it cannot permanently remove the trapped charge (e.g., trapping may recur when the voltage goes back to zero), then the $R$ and $C$ decrements are volatile. On the other hand, if electron removal is further aided by the work-function differential between the two electrodes, then it can be shown that permanent removal of a trapped electron is possible for one voltage-biasing under one polarity but not the other[33]. This is what we have: In negative biasing, the metallic state obtained is non-volatile; in positive biasing, it is not.

In summary, like most inorganic glasses macroscopic amorphous Si and Ge are insulators, but we have found in ~10 nm films of doped/undoped amorphous Si and Ge a nested exotic metal with an unexpectedly heavy electron mass and uncommonly strong electron-phonon coupling. These remarkable features stem from the extreme local structures of amorphous Si and Ge, which are further exacerbated by O or N doping to enable stimulus-triggered but temperature/time-insensitive metal-insulator transitions. In such metal, there is little electron hopping between localized states, or else the inelastic hopping process would have destroyed electron coherence, hence the interference phenomena so prominently displayed here. The picture also applies to other nanostructured amorphous insulators—or metals, rather. They likely include random insulator-metal "alloys" such as $Si_{0.95}N_{1.33}(Pt,Cr)_{0.05}$ and random Mott insulators such as $HfO_2$ and $Al_2O_3$[30,34], for they too exhibit thickness-mediated, pressure-responsive nanometallic transitions. More broadly, we believe ultrasoft local structures with a negative $U$ are a structural part of amorphous inorganic insulators, even though their profound effects—on coherent electron's effective mass at low temperature and on electron trapping at all temperature—are only unmasked here for the first time thanks to the discovery of nanometallicity. These nanomaterials are suitable for non-volatile memory[31], but other applications are possible. For example, amorphous nanometallic Si may incorporate additional electro-active elements to form nanostructured hybrid electrodes for electrochemical energy cells.

**METHODS**

**Materials Fabrication.** Amorphous thin films of Si and Ge with and without dopants were sputter deposited on two types of substrates: a thermal-oxide-coated 100 p-type silicon single crystal, and a two-side polished fused silica substrate. Before deposition, the unheated substrate was first coated by a 30 nm thick Mo bottom electrode using DC sputtering, and the Si or Ge film was next deposited by RF sputtering without breaking the vacuum. Targets used were Si (bulk resistivity > 1 Ohm-cm), n-type Si (P-doped, bulk resistivity < 0.1 Ohm-cm), p-type Si (B-doped, bulk resistivity ~ 0.005-0.020 Ohm-cm) and Ge (Sb-doped, bulk resistivity ~ 5-40 ohm-cm). When desired, O/N was incorporated during Si/Ge sputtering by either injecting $O_2$/$N_2$ gas together with Ar gas into the sputtering chamber, or RF co-sputtering an oxide/nitride target chosen from $Si_3N_4$, AlN, $SiO_2$, $Al_2O_3$ and $HfO_2$. The electrical properties of films prepared by these different doping methods were similar (see, e.g., **Extended Data Fig. 5**) so we mostly used $Si_3N_4$ for N incorporation and $Al_2O_3$ for O incorporation. Composition can be tuned by the flow rate of $O_2$/$N_2$ gas or by the sputtering power of the oxides/nitride target. Finally, a 40 nm thick Pt top electrode was deposited by RF sputtering either through a shadow mask that defined cells of 50-250 μm in radius, or onto a lithography-defined pattern of 1-20 μm radius cells followed by a lift-off process. The former type of cells was used for most transport measurements and hydraulic pressure experiments, while the latter type was used for most magnetic pressure experiments.

**Materials Characterization.** X-ray diffraction (XRD) patterns of Si/Ge films were measured by Rigaku GiegerFlex D/Max-B diffractometer. The patterns were compared with that of a bare single crystalline 100-Si substrate, and showed no additional peak other than the 100-Si peak of the substrate (see **Extended Data Fig. 1**). To determine the film thickness, atomic force microscopy (AFM, Asylum MFP-3D) was used to measuring the step height created by deposition when one side was intentionally blocked. X-ray photoelectron spectroscopy (XPS) on Si films doped with O or N was collected using a RBD upgraded PHI-5000C ESCA system (Perkin Elmer) with Mg $K_α$ radiation ($hv$ = 1253.6 eV). Binding energies were calibrated by referring to carbon (C 1 $s$ = 284.6 eV). Composition of O and N in Si films was then determined using Si:2$p$, O:1$s$, and N:1$s$ peaks by comparing them to calibrated signals from standard SiO$_2$ and Si$_3$N$_4$ samples.

**Electrical Measurement Sample Preparation.** Temperature dependent electrical properties of films with and without a magnetic field were measured in Quantum Design Physical Property Measurement System (PPMS, 2-300K, −9-9 T) at Penn's Shared Experimental Facilities and two superconducting magnets at National High Magnetic Field Laboratory (SCM2, 0.3-2K, −18-18 T; and SCM1, 0.018-2K, −18-18 T). Si or Ge films of different thickness or preset in different resistance states were mounted on sample holders, either a sample puck for PPMS or a set of 16-pin dip sockets for SCM1 and SCM2. With one end soldered to the pins on the sample holder, a gold wire was silver-paste-bonded to the top Pt electrode of a Si/Ge cell of a radius of 250 μm. Another wire was similarly bonded to one edge of the bottom Mo electrode to form a two-terminal connection to the cell. A three-terminal connection was formed by further bonding a third gold wire to the opposite edge of the Mo electrode; these three-terminal connections used samples deposited on a fused silica substrate, which is insulating.

**Electrical Measurement Setup.** During measurements, the following setups were used. (i) Two-point DC resistance (defined as $V_{2pt}/I$) was obtained using a Keithley 237 High Voltage Source Measure Unit to apply a constant voltage of 0.01 V ($V_{2pt}$) across the top and bottom electrodes, while measuring the current ($I$) passing through the films. (ii) Three-point DC resistance (defined as $V_{3pt,\ DC}/I$) was measured using a Keithley 237 to apply a constant current of 100 μA ($I$) between one lead of the top electrode and one edge of the bottom electrode, while measuring the voltage using a Keithley 2182A nanovoltmeter between the other lead of the top electrode and the other edge of the bottom electrode. This configuration removes the spreading resistance but still include the resistance (as well as the interface resistance) of the top and bottom electrodes within the cell area. (see Ref. 22 for details of further estimating the resistance of this part.) (iii) Three-point AC resistance (defined as $V_{3pt, AC}/I$) was measured using an SR 830 lock-in amplifier and a standard resistor of 100 kΩ. The amplifier sent a sine wave of 1 V amplitude at 31 Hz to the standard resistor and the cell, in serial connection. Since the resistance of our cell in the metallic state is typically much less than 100 kΩ, the above is equivalent to applying a constant current of 10 μA ($I$) across one lead of the top electrode and one side of the bottom electrode. Meanwhile, using the lock-in amplifier, the voltage (A-B voltage or $V_{3pt, AC}$) between the other lead (A) of the top electrode and the other side of the bottom electrode (B) was measured when locked to the same frequency of 31 Hz at a time constant of 1 s. An auto offset and 10× signal expansion were used to improve the voltage resolution to enable measuring small resistance changes. (i)-(ii) were used in PPMS runs, and (ii)-(iii) were used in SCM1 and SCM2 runs. All the measurements were performed in the Ohmic regime (as shown in **Extended Data Fig. 2**) and yielded almost identical results when fitted by the same set of parameters (e.g., (ii) was used in **Fig. 2b-c** and (iii) used in **Fig. 2d-e**).

**Temperature/Field Sweep.** Electrical measurements were mainly conducted under two temperature/field-sweep conditions: cooling/heating at a fixed magnetic field (often zero-field, see **Fig. 1a**, **Fig. 2a-b**, **Fig. 2e**, **Extended Data Fig. 2c-f**), and ramping magnetic field at fixed temperature (**Fig. 2c-d**, and **Extended Data Fig. 3** and **Fig. 4b-c**). During these measurements, synchronized voltage, current, temperature and field data were recorded while the heating/cooling rate was set at an appropriate value. Sweeping of the magnetic field was typically at 0.5 T/min in PPMS and 0.3 T/min in SCM1 and SCM2. The field/sample-orientation dependence of magnetoresistance was determined in SCM1 and SCM2 by rotating the sample at 3 degrees/min in a fixed magnetic field (**Extended Data Fig. 3k**), or by rotating the sample to a new orientation, then sweeping the field to ±18T (**Extended Data Fig. 3i-j**). Precaution against temperature/magnetic-field/orientation/voltage hysteresis was taken by repeating all the temperature/magnetic-field/orientation/voltage sweeps in two directions to ascertain their ramping curves overlap exactly. For the same temperature/field range, data collected in PPMS, SCM1 and SCM2 were in excellent agreement with each other despite the fact that they were measured several months apart. Thus, our materials and sample cells are apparently highly uniform, stable and reproducible.

**Data Analysis.** Despite different film thickness, resistance and composition, all our PPMS, SCM1 and SCM2 measurements used sample cells of the same size and configuration, which are likely to have the same parasitic load resistance due to leads, electrodes, spreading resistance and electrode/film interfaces. Moreover, although the measured resistance of necessarily contains some parasitic load resistance, we found it possible to precisely determine the resistance change of the films in our sample cells as illustrated by the following example. Consider two cells that contain two different films that are self-similar in the following sense: Both films have a magnetoresistance at 10 T that is 1% of its 0 T resistance. (Note that their cell magnetoresistance, which includes the magnetoresistance of both the film and the parasitic resistance, is not self-similar even if film's magnetoresistance is.) If so, then the difference of the two cell-resistances will also show a magnetoresistance of 1% of its 0 T value, because unlike the cell-resistance the difference does not include the parasitic resistance. This example suggests that, instead of analyzing the two-terminal cell resistance individually, we can analyze a set of them, and compare their pair differences, which will carry the same information of the relative resistance change of the film. (There is an analogous case in diffraction of matter: The diffraction intensity is the Fourier transform of the vectors of atom-pairs; it is *not* the Fourier transform of the position vectors of atoms. Nevertheless, the diffraction intensity still informs the structure of matter satisfactorily.) We have formalized this analysis and reported it elsewhere[20]. Below is a summary of the procedure used for data analysis here.

Let the resistances of two self-similar cells, 1 and 2, increase from $R(1)$ and $R(2)$ at $(T,B)=(0,0)$ to $R'(1)$ to $R'(2)$ at $(T,B)$, respectively. Then the relative conductivity change $\Delta\sigma/\sigma$ of the can be obtained from the relative change of the resistance difference of the cell pair, $-\Delta R/R_0$. Here, $R_0 = R'(1) - R'(2)$ and $\Delta R = (R'(1) - R'(2)) - (R(1) - R(2)) = (R'(1) - R(1)) - (R'(2) - R(2))$ is the change of the resistance difference due to the change of $(T,B)$. The above result is exact. This method can be used to accurately determine $\Delta\sigma/\sigma_0 = -\Delta R/R_0$ down to 0.1%.

In our data analysis, for each composition we first selected a set of metallic cells that differ in either Si/Ge thickness or the resistance value. These cells will be named cell 1, 2, etc. Assuming they are self-similar, we followed the resistance difference method to calculate $-\Delta R$. (The resistance difference between two cells used for the following analysis was in the Ohmic regime.) The self-similarity assumption is considered validated if (i) the resultant $-\Delta R$ obeys a scaling law; for example, $-\Delta R$ of various $(T,0)$ follows a temperature scaling law, and (ii) $\Delta\sigma/\sigma = -\Delta R/R$ of all the pairs falls on a universal curve/line consistent with the scaling law. Comparing thus obtained $\Delta\sigma/\sigma_0 = -\Delta R/R_0$ data with Eq. (1) then allows determination of the model parameters for each composition. Since there are only two parameters, $\alpha/\sigma_0$ and $\tilde{F}_\sigma$, in Eq. (1), two independent sets of data, e.g., temperature dependence and magnetoresistance, will suffice for solving the parameters. A final check of self-similarity was made by using the same values of α/σ₀ and $\tilde{F}_\sigma$ for all the films of the same composition—but of different thickness and/or resistance—to see if an excellent agreement between the Eq. (1) prediction and the experimental data can still be obtained. This is the case as illustrated by the overlapping data points and the theoretical solid curves in **Fig. 2**, **Extended Data Fig. 3** and **Fig. 4**.

**Calculating Electron Properties from Model Parameters.** As described in the main text, values of $F_\sigma$, $\sigma_o$ and $D$ can be obtained by model fitting of the experimental data. They can next be used to calculate diffusive electron properties if we assume $l$=0.27 nm and a spherical Fermi surface, and identified $\sigma_o$ with $\sigma_D$, as follows. First, $v_f$ and τ can be calculated using $v_f = 3D/l$ and $\tau = l/v_f$. Next, with a Fermi sphere and $\sigma_o = \sigma_D$, which can be written as $\sigma_D = \dfrac{ne^2\tau}{m^*}$, we can replace $n$ by $\dfrac{k_f^3}{3\pi^2} = \dfrac{(m^*v_f)^3}{3\pi^2\hbar^3}$ and τ by $l/v_f$, so $\sigma_o = \dfrac{ne^2\tau}{m^*} = \dfrac{(m^*v_f)^2 e^2 l}{3\pi^2\hbar^3} = \dfrac{e^2 m_e^2}{3\pi^2\hbar^3}\left(\dfrac{m^*}{m_e}\right)^2 v_f^2 l$. Thus, the effective mass $m^*/m_e$ can be calculated by $\left(\dfrac{3\pi^2\hbar^3}{e^2 m_e^2}\dfrac{\sigma_o}{v_f^2 l}\right)^{1/2} = 0.013\dfrac{\sqrt{\sigma_o l}}{D}$, all quantities in SI units.

Finally, $k_f = \dfrac{m_e}{\hbar}\dfrac{m^*}{m_e}v_f = 8635\dfrac{m^*}{m_e}v_f$ and $n = \dfrac{k_f^3}{3\pi^2}$, again in SI units. These calculated properties are listed in **Extended Data Table 1**.

**Voltage Triggered Metal-Insulator Transition.** Voltage was applied to O/N-doped Si/Ge films to transition them between the insulating state and multiple metallic states. In a typical setting, samples were placed in a probe station, and a bias voltage supplied by a Keithley 237 was applied to the top Pt electrode while the bottom electrode was grounded. A positive bias of ~1 V triggers the transition from a metallic state to the insulator state, whereas a negative bias of ~ −1 V triggers the opposite transition to various metallic states—their access determined by the current compliance used. Voltage-triggered transitions were similarly performed in PPMS and SCM 2, again using a Keithley 237 to sweep the DC voltage with an appropriate current compliance to prevent a sudden temperature burst due to Joule heating by the transition current. These experiments were performed down to 2K; transitions at lower temperature were not attempted because of difficulty to ascertain the transition temperature due to Joule heating. Because the resistance of metallic Si/Ge films can be arbitrarily small, the actual threshold transition voltage $V^*$ of ±1 V was calculated by $V^*=V_s\times(R_{cell}-R_{BE})/R_{cell}$, where $V_s$ is the voltage applied to the top electrode at the time of transition and $R_{BE}$ ~500 Ω for the parasitic resistance primarily due to the spreading resistance of the bottom electrode, which can be separately determined by linear regression of $1/R_{cell}$ on $1/V_s$ as discussed in details elsewhere[33]. This procedure provides the electrical transition curves shown in **Fig. 3a**, **Extended Data Fig. 5**, **Fig. 7a-b**, and **Fig. b,** which report the actual voltage $V_f$ on the films ($V_f=V_s\times(R_{cell}-R_{BE})/R_{cell}$). They reveal a sudden increase of voltage during the +1 V transition to insulator state and a zigzag shape of voltage around − 1 V during the transition to metallic state, two features also seen elsewhere in other switching materials[33].

**Pressure Triggered Insulator-to-Metal Transition.** A hydraulic pressure was used to cause insulator to metal transition. Before the pressure treatment, the two-point resistance of each Si/Ge cell in the cell array on the same chip was read at 0.1 V or pre-transitioned to certain resistance state using Keithley 237. The transitioned states include both the insulator state and the metallic state, while other cells were left in their current states after reading the resistance.

Next, the chip was disconnected from the voltage source, wrapped in an aluminum foil, vacuum-sealed in an elastomer bag, and suspended in a liquid-filled pressure vessel (Autoclave Engineers, Erie, US) that was charged to a preset hydraulic of 2-350 MPa at room temperature and held for <5 min before sample removal. The resistance of each cell was read again at 0.1 V and compared with its pre-pressure-treatment value, and the result is presented in a cumulative probability curve such as **Fig. 3b**. Some higher pressure (up to 1 GPa) experiments were also similarly performed in a hydraulic pressure vessel (Dr. CHEF) at Takasago Works, Kobe Steel, at Takasago, Japan.

**Magnetic-Pressure Triggered Insulator-to-Metal Transition.** A magnetic pressure burst was used to trigger insulator-to-metal transition. In principle, a magnetic pressure may be generated by passing a burst magnetic flux between the Si/Ge-filled gap of the two electrodes, which form a metallic "container" that confines the burst magnetic field. (At high frequency, the gap resistance at the edges of the electrodes is very small so the two electrodes form a continuous circuit.) Assuming Si has a relative permeability of unity, the magnetic pressure is $(B/0.501)^2$ with the pressure expressed in bar (1 bar=0.1 MPa) and $B$ in T. The burst magnetic flux was received from an electron bunch, which is a spatially localized bundle of 20 GeV electrons generated at Stanford Linear Accelerator Center (SLAC) using the FACET facility. Each bunch contained ~$10^9$ electrons (>1 nC) that are narrowly collimated (~25 μm). It had a short duration, passing in ~0.1 ps, which is the time for the bunch to travel 25 μm at near the speed of light, and was available on a bunch-by-bunch basis. We only allowed each cell to see one bunch during the experiment; after each shot the sample was moved to a new position before the second shot was fired. The electron bunch hit the sample chip in the normal direction. Since the maximum magnetic field around the bunch is ~ 65 T at the edge of the bunch, i.e., ~ 25 μm from the flight path, and it decays with the radial distance *r* from the bunch roughly according to 1/*r*, we can estimate the magnetic pressure in each cell from the cell location relative to the flight path. (To maximize the induced magnetic pressure inside the cell, we chose the cell size to be 20 μm, comparable to the bunch size.) So, the estimated maximum magnetic pressure is ~1,680 MPa, and at 500 μm away it decays to ~4.2 MPa. (The above estimates are lower bounds since they do not consider the pressure caused by the induced current in the electrodes.[15]) Because of symmetry, however, there is no magnetic field at the center of the bunch. This was verified by the observation in some experiments one or two center cell that that suffered neither physical damage nor resistance change. (It also provided direct evidence that the impact damage due to momentum transfer was minimum, which is expected because the collision cross-section of a relativistic electron at 20 GeV is very small. See further discussion in Ref. 33) The electric field is radial and follows the same radial variation as the magnetic field, but it is unimportant for this experiment as previously established in Ref. 15.

Before the magnetic-pressure treatment, cells were pre-transitioned by a voltage to certain resistance states, with their two-point resistance values recorded at 0.1 V by a Keithley 237. Their resistance was again read in the same way after the magnetic-pressure treatment and compared with the pre-treatment value. In a typical representation of the data, each cell is colored to indicate its resistance value before and after the treatment, and the colored maps are presented to aid comparison. In some experiments, the chip was covered by a photoresist, which is not a conductor and has no effect on the magnetic field. But because it can be blown away by a large magnetic field, it serves as a marker to help identify the flight path of the electron bunch. Since the cell size is about the same as the bunch size, maps that have hundreds of cells each appearing as a "dot" with changed colors (**Fig. 3c, Extended Data Fig. 7c-d**) provide direct evidence for the far-field effect of an electron bunch.

**Capacitance and AC impedance.** Capacitance measurements were performed using a HP4192A impedance analyzer ($10^2$-$10^7$ Hz, 50 mV oscillation level) under different DC bias. Capacitance values of Si/Ge films in different resistance were extracted either by direct measurement that presumes a serial *R-C* circuit at 100 kHz or by fitting the impedance spectra from 100 Hz to $10^7$ Hz using an equivalent circuit or directly from the resonance frequency. Within the same Si/Ge cells, resistance values that differ by orders of magnitude can be obtained by applying a negative voltage to preset resistance states before capacitance measurements (for O/N doped Si/Ge films) or by applying a positive DC bias during capacitance measurements (for pure Si/Ge films or Si/Ge lightly doped with O/N that are unable to transition but have non-linear resistance with respect to voltage, such as the one in **Fig. 3a**).

**Acknowledgements**

This research was supported by the US National Science Foundation Grant No. DMR-1409114. The use of facilities including PPMS at Penn's LRSM supported by DMR-1120901 is gratefully acknowledged. Work was also performed at the National High Magnetic Field Laboratory (NHMFL), which is supported by National Science Foundation Cooperative Agreement No. DMR-1157490 and the State of Florida, at SLAC National Laboratory supported by the US Department of Energy, Office of Basic Energy Sciences, and at Takasago Works, Kobe Steel, at Takasago, Japan. Experimental assistance of Prof. Jay Kikkawa (Penn), Drs. Ju-Hyun Park (NHMFL), Hongwoo Baek (NHMFL) and Ioan Tudosa (SLAC) is gratefully acknowledged.

**Author Contributions**

**Author Information**

Reprints and permissions information is available at www.nature.com/reprints. The authors declare no competing financial interests. Correspondence and requests for materials should be addressed to I.W.C. (iweichen@seas.upenn.edu).

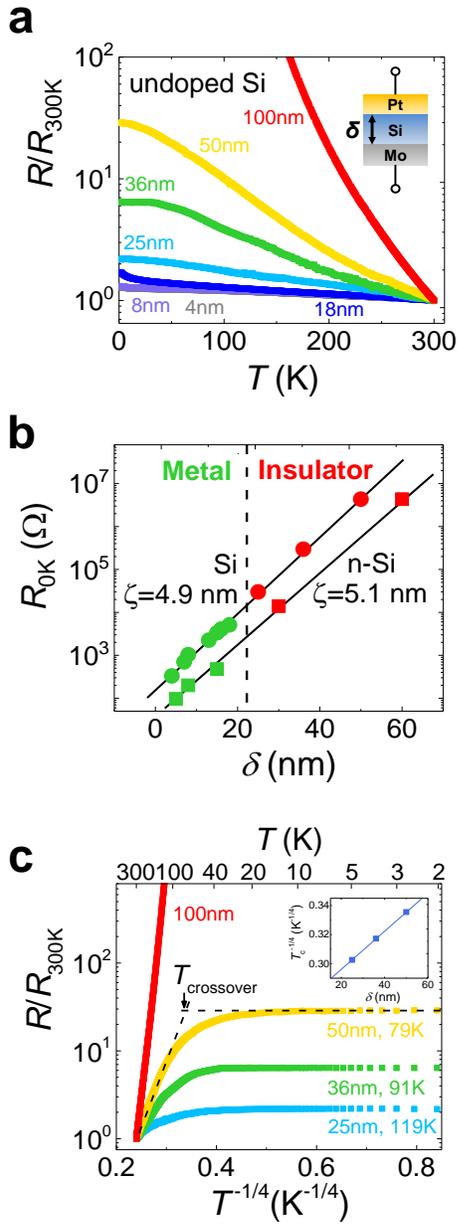

**Figure 1 | Thickness-mediated insulator-metal transition in amorphous Si films. a,** Thickness-dependent two-point Ohmic resistance $R$ (normalized by $R_{300K}$) across the Pt/Si/Mo stack (inset) *vs.* temperature $T$. Insulator's (>25 nm) resistance changes by orders of magnitude while metal's (<25 nm) is nearly constant. Resistance increases at $T\sim 0K$ in 18 nm has same origin as resistance minimum in **Fig. 2a**. **b,** Extrapolated 0K resistance follows $\exp(\delta/\zeta)$ dependence. Circles: Si; squares: n-type Si. **c,** Normalized resistance *vs.* $T^{1/4}$ for insulating films obeys "1/4" law at high temperature but saturates below $T_{\text{crossover}}$ (marked), which depends on film thickness $\delta$ as predicted by theory of variable-range-hopping. Temperature marked on the top.

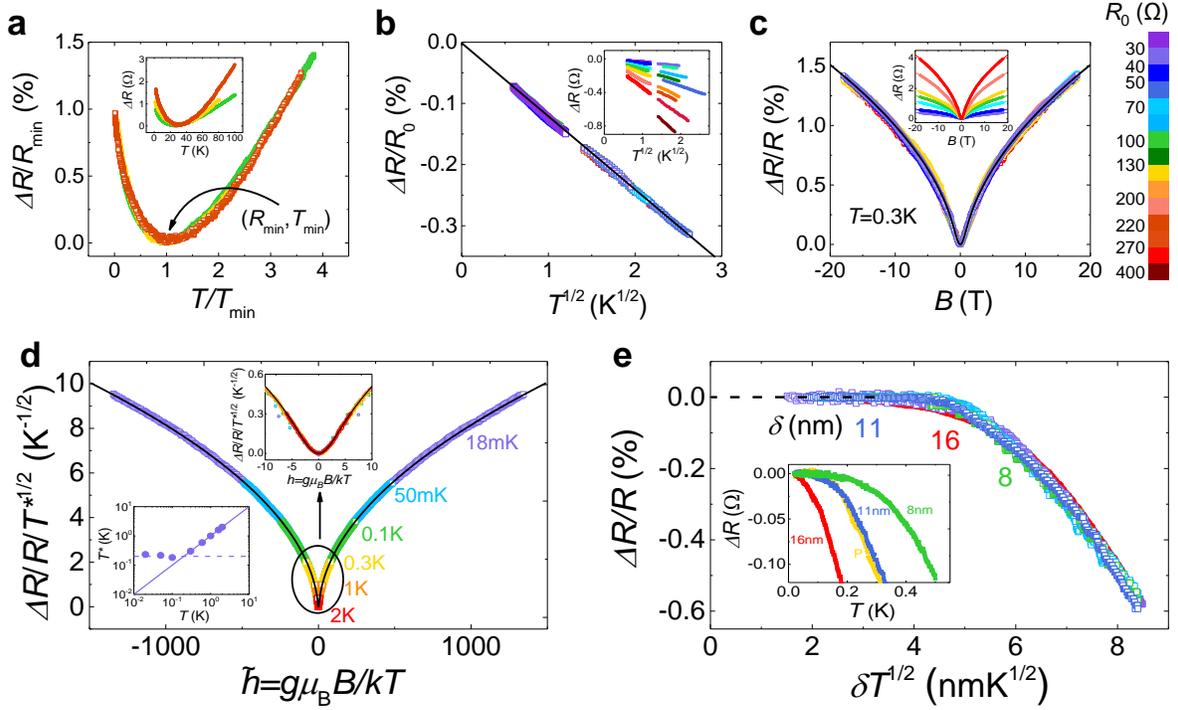

**Figure 2 | Weak localization dominated by 3D electron-electron interaction causing resistance changes on (*T*,*B*) perturbation in metallic n-Si:0.45N films. a,** Resistance minimum ($R_{min}$) at $T_{min}$ for 3 metallic states in 11 nm film with overlapping normalized plots, $\Delta R/R_{min}$ vs. $T/T_{min}$. Inset: Unnormalized data with same $T_{min}$~30K. **b,** Relative resistance changes $\Delta R/R_0$ of 16 metallic states in 11 nm film with overlapping $T^{1/2}$ dependence below $T_{min}$ follows prediction of Eq. (1) shown as solid line. Inset: Unnormalized data of 16 states with $R_0$ indicated by color spectrum on far right. **c,** Relative magnetoresistance (0.3K) of 7 metallic states in 11 nm film with overlapping plots follows prediction of Eq. (1) shown as solid curve. Inset: Unnormalized data with $R_0$ indicated by color spectrum on far right follows predictions of Eq. (1) shown as solid curves. **d,** Relative temperature-dependent magnetoresistance divided by $T^{1/2}$—or by $T^{*1/2}$ shown in lower inset at <0.2K—with overlapping $g\mu_B B/kT$ dependence follows prediction of Eq. (1) shown as solid curve with $B^{1/2}$ dependence at high field. Upper inset: Enlarged view of small $g\mu_B B/kT$ data with $B^2$ dependence predicted by Eq. (1). **e,** Relative resistance changes of 3 film thickness (filled symbols)—one (11 nm) in 3 metallic states (open symbols)—with overlapping $\delta T^{1/2}$ dependence saturating (approaching dashed line) below $\delta T^{1/2}$~4.8 (nmK$^{1/2}$). Inset: Unnormalized data of different thickness. Another pressure-transitioned metallic state in 11 nm film (yellow stars marked by "P") has almost identical saturation behavior as voltage-transitioned metallic state (blue marked by "11 nm"). In **b-d**, predicted curves by Eq. (1) all use $\tilde{F}_\sigma$=0.75 and $\alpha/\sigma_o$=5.76×10$^{-3}$.

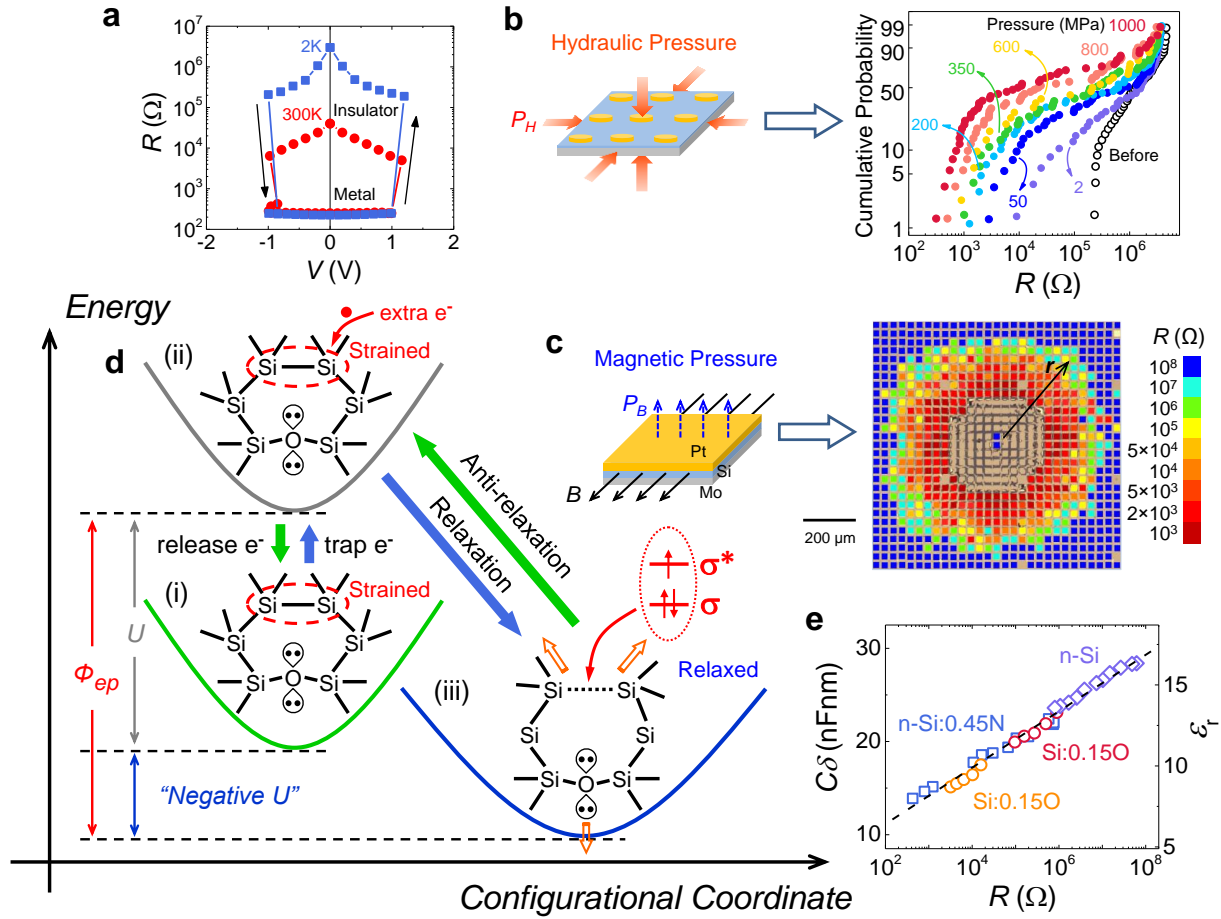

**Figure 3 | Trapped-electron regulated, electromechano-stimulus-triggered metal-insulator transition in amorphous Si films. a,** Reversible transitions of 11 nm n-Si:0.45N film between insulator state (higher resistance at 2K than at 300K) and metal state (same resistance at 2K and 300K) triggered by comparable voltage ~ ±1 V at 2K and 300K. Arrows indicate transition direction. **b,** Uniform hydraulic pressure $P_H$ (see schematic on left) applied to Si:0.48O cells lowers cell resistance from initial resistance distribution (shown as cumulative probability and marked as "Before") to new distributions. The distribution shifts to the left when $P_H$ increases from 2 MPa to 1,000 MPa. **c,** One electron bunch (20 μm dia.) hitting $r=0$ generates one $10^{-13}$ s pulse of magnetic field $B\sim1/r$ (roughly) and magnetic pressure $P_B \sim 1/r^2$ (see schematic on left) inside n-Si:0.45N cells, which lowers cell resistance (all preset to ~100 MΩ, corresponding to dark blue in outer cells) to values indicated by new cell colors according to spectrum on right. Each "dot" in figure on right is a 20 μm cell. Excessive $P_B$ generates excessive biaxial tension in top electrode and ruptures it (see uncolored earth-tone cells at small $r$.) Because $B=0$ at center, two center cells remain intact with initial resistance unperturbed. Lower bound estimate for $P_B\sim1,680$ MPa at $r=20$ μm (where $B$ peaks at edge of electron bunch), ~90 MPa at $r=180$ μm (where electrode rupture ends), and 20 MPa at $r=390$ μm (where transition ends). **d,** Energy-configuration coordinate diagram specialized to a strained Si-Si bond next to Si-O-Si: (i) Initial

stable state (strained Si-Si bond); (ii) metastable state with positive Hubbard-$U$ (extra electron entering the strained Si-Si region); (iii) stable state (Si-Si triple bond with two bonding electrons and one anti-bonding electron; (arrowed) relaxation caused by longer bond length and Si electron repulsion against lone-pair electrons of O) with negative $U$ aided by $\phi_{ep}$ due to relaxation. Voltage can deliver extra electron to trapped site and, later, destabilize the trapped electron. Pressure cannot deliver extra electron but can also trigger detrapping by enabling anti-relaxation returning (iii) to (ii). Stably trapped electron in (iii) exerts long-range repulsion against itinerant electrons. **e,** Capacitance $C$ of Si films normalized by thickness $\delta$ correlates with resistance $R$; different $R$ values obtained by setting appropriate voltage on cell—guided by transition curves such as the 300K one in **a**. Right: Converted relative dielectric constant $\varepsilon_r = C\delta/\varepsilon_0 A$.

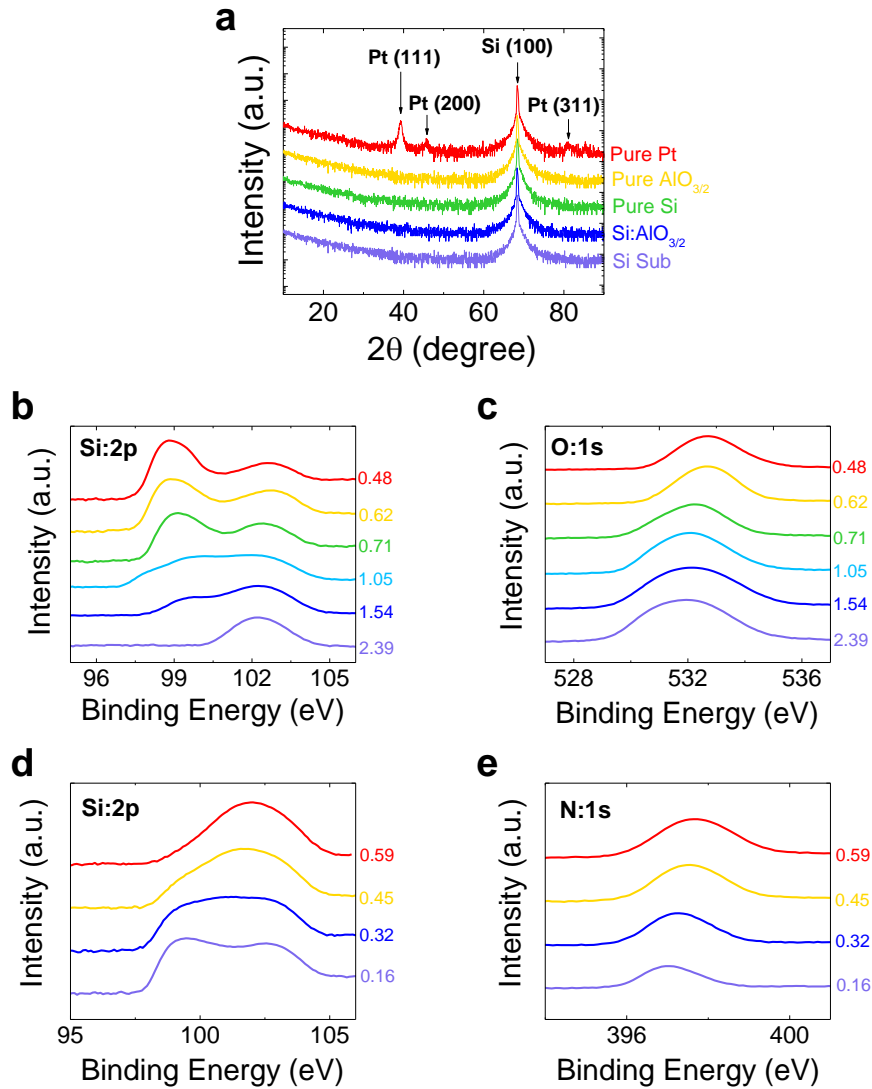

**Extended Data Figure 1 | Material characterization of Si thin films. a,** XRD patterns of thin films of sputtered Pt, $AlO_{3/2}$, Si, and co-sputtered Si:$AlO_{3/2}$ (giving Si+O) and their single-crystalline 100-Si substrate. No additional peak other than those of the substrate appears in Si, $AlO_{3/2}$, or their mixture. Pt has additional peaks of Pt-111, 200 and 311. Intensity in logarithmic scale aids detection of trace phases. **b-c,** XPS spectra of Si:2*p* (**b**) and O:1*s* (**c**) in Si+O films. Numbers on the right indicate the content *x* of O as in $SiO_x$ determined by comparison with $SiO_2$ standard. **d-e,** XPS spectra of Si:2*p* (**d**) and N:1*s* (**e**) in Si+N films. Numbers on the right indicate the composition *y* of N as in $SiN_y$ determined by comparison with $Si_3N_4$ standard.

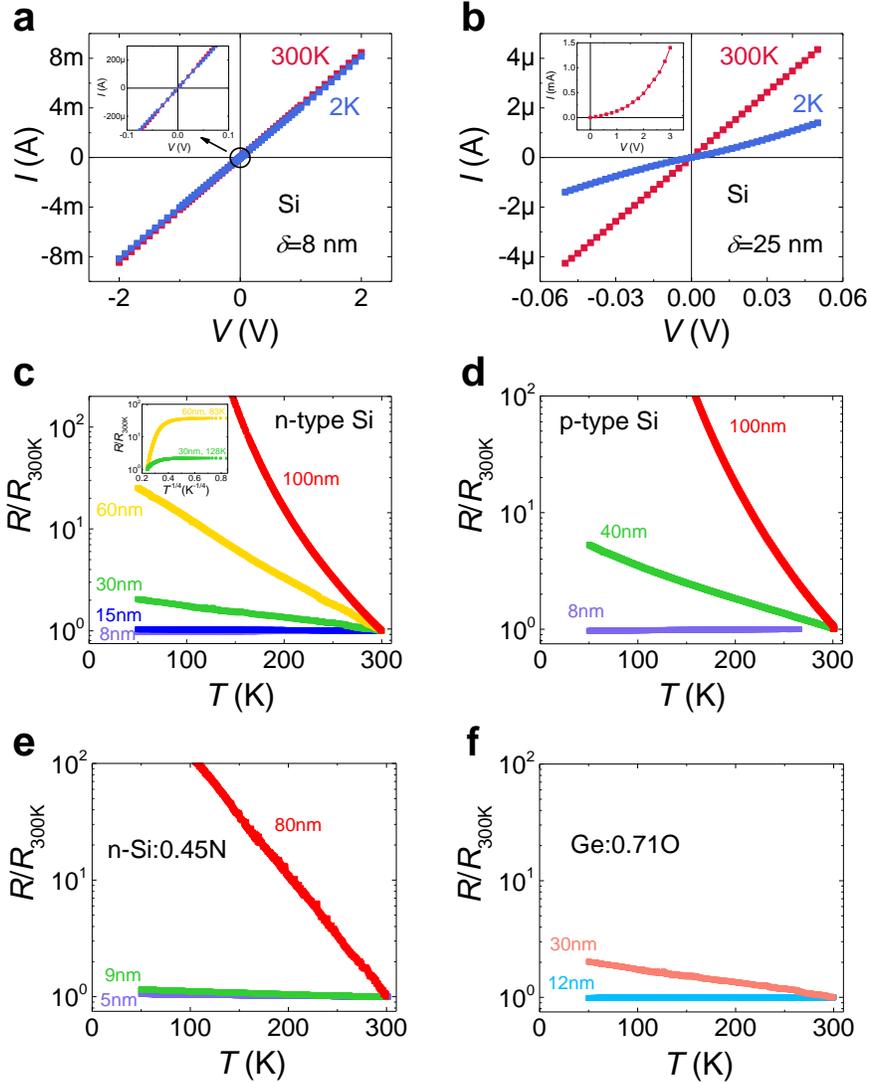

**Extended Data Figure 2 | *I-V* curves and resistance ratio identify thickness-triggered insulator-metal transition in various amorphous Si and Ge films. a,** Ohmic *I-V* curves of metallic 8 nm Si film in the voltage range of − 2 V - 2 V at 300K and 2K. Inset: Enlarged view from − 0.1 V to 0.1 V. **b,** Ohmic *I-V* curves at 300 K and 2 K of insulating 25 nm Si film at small voltage, below 0.02 V, vs non-Ohmic one at higher voltage at 300K in the inset. **c-f**, Thickness-triggered insulator-to-metal transitions in n-type Si (**c**); p-type Si (**d**); n-Si:0.45N (**e**); and Ge:0.71O (**f**). Resistance measured in Ohmic regime. Inset of **c**: *R* follows $T^{1/4}$ until saturation at low temperature in insulating n-Si films, which resembles **Fig. 1b** in main text.

Please remove "undoped" from a-b.

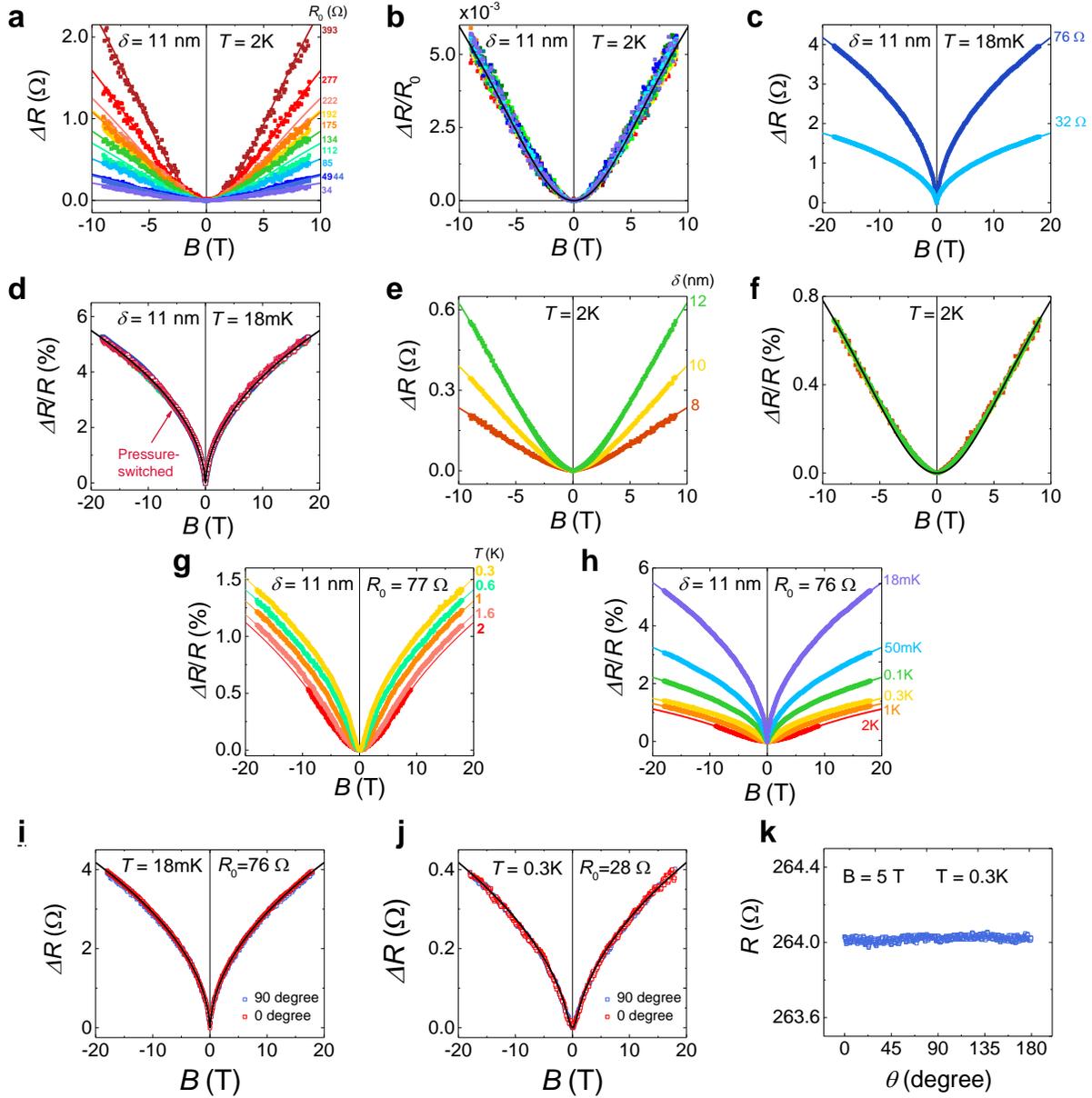

**Extended Data Figure 3 | Relative resistance changes of metallic n-Si:0.45N films caused by ($T,B$) perturbation is independent of metallic states, thickness, temperatures and orientations, and agree with Eq. (1) predictions (solid curves in a-k) with $\tilde{F}_\sigma = 0.75$ and $\alpha/\sigma_o = 5.76 \times 10^{-3}$. a,** Magnetoresistance ($\Delta R$), at 2K, of 11 nm film in 10 metallic states. Numbers on the right indicate the extrapolated 0K resistance $R_0$. **b,** Data of all states in **a** overlap when plotted as as relative magnetoresistance ($\Delta R/R_0$). **c,** Magnetoresistance, at 18mK, of 11 nm film in 2 metallic states. **d,** Data of both states (two shades in blue) in **c**, together with pressure-transitioned metallic states (two shades in red) of the same sample, overlap when plotted in relative magnetoresistance. **e,** Magnetoresistance, at 2K, of 3 films of different thickness. **f,** Data of 3 films in **e** overlap when plotted in relative magnetoresistance. **g,** Relative magnetoresistance of another state ($R_0$=77 Ω) of 11 nm film at various temperatures from 0.3K to 2K. **h,** Relative magnetoresistance of one state ($R_0$=76 Ω) of 11 nm film at various temperatures from 18mK to 2K. (Same data as shown in **Fig. 2d** in main text.) **i,** Magnetoresistance, at 18mK, in two sample orientations overlap. 0 degree: field parallel to film surface; 90 degree: field normal to film surface. **j,** Same as **i**, for another resistance state, at 0.3K. **k,** Orientation-independent magnetoresistance of 11 nm film at 5 T and 0.3K.

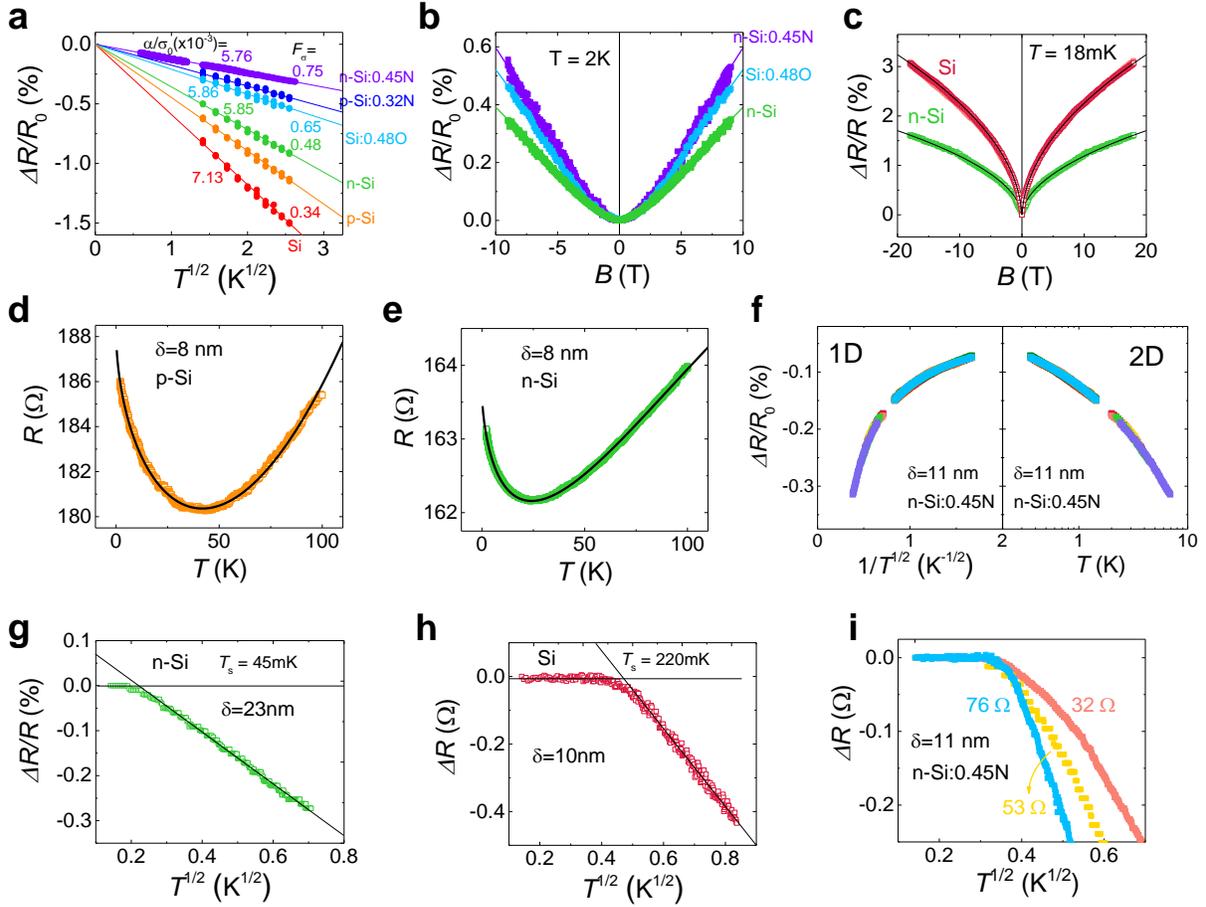

**Extended Data Figure 4 | Resistance change and saturation of metallic Si films caused by ($T,B$) perturbation depend on composition. a,** Relative resistance changes $\Delta R/R_0$ of metallic Si, p-Si, n-Si, Si:0.48O, p-Si:0.32N and n-Si:45N have different slopes against $T^{1/2}$. Each plot of first three compositions includes data of different film thickness, while each plot of the last three compositions includes data of different metallic states. Solid lines are Eq. (1) predictions with material parameters ($\alpha/\sigma_0$ and $F_\sigma$) listed next to lines. **b-c,** Relative magnetoresistance of n-Si, Si:0.48O and n-Si:0.45N at 2K (**b**) and n-Si and Si films at 18mK (**c**) follows Eq. (1) predictions (solid curves) with the same material parameters listed in **a**. Each plot contains data from several different states/thickness. **d-e,** Resistance minima in 8 nm p-Si film (**d**) and 8 nm n-Si film (**e**), curve-fitted by $R_0+AT+BT^{p/2}+CT^{1/2}$ ($A>0$, $B<0$, $C<0$, $p=1.5$, with $B$ fixed as it is attributed to the electrode component that is presumably the same for all films.) **f,** Relative resistance change $\Delta R/R_0$ of 11 nm n-Si:0.45N film in 20 different states does not follow $1/T^{1/2}$ dependence (left, the 1D prediction) or $\log T$ dependence (right, the 2D prediction). This should be compared with $T^{1/2}$ dependence, which it follows as shown in **Fig. 2b**. **g-h,** Resistance saturates at low temperature in 20 nm n-Si film (**g**) and 10 nm Si film (**h**), from which diffusivity $D$ can be calculated. **i,** Resistance of different metallic states in 11 nm n-Si:0.45N film saturates at about the same temperature, indicating same dephasing length in different states. When plotted as relative change $\Delta R/R_0$, these data overlap with other saturation curves of films of the same composition but in different thickness. See **Fig. 2e** where these data are shown as open symbols.

**Extended Data Table 1 | Properties of coherent conducting electrons in amorphous silicon nanostructures.** Values of $F_\sigma$, $\sigma_0$, and $D$ were obtained from data fitting to $T^{1/2}$ resistance dependence (**Fig. 2b**, **Extended Data Fig. 4a**), magnetoresistance (**Fig. 2c-d**, **Extended Data Fig. 3** and **Extended Data Fig. 4b-c**), and resistance saturation (**Fig. 2d**, **Extended Data Fig. 4g-h**). From the above properties, values of $v_f$, $\tau$, $m^*/m_e$, $k_f$, and $n$ were next calculated by setting $l=0.27$ nm and $\sigma_0$ to Drude conductivity (see Method). Small $D$ and $v_f$ are due to the heavy effective mass caused by a large local $\phi_{ep}$. Additional values of $F_\sigma$ and $\sigma_0 D^{1/2}$ (~$\sigma_0/\alpha$) obtained from data fitting to $T^{1/2}$ resistance dependence and magnetoresistance also shown for another material.

| Si-type | $F_\sigma$ | $\sigma_0 D^{1/2}$ $(\Omega s^{1/2})^{-1}$ | $D$ $(cm^2/s)$ | $\sigma_0$ $(\Omega cm)^{-1}$ | $l$ (nm) | $v_f$ (cm/s) | $\tau$ $(10^{-14}s)$ | $m^*/m_e$ | $k_f$ $(nm)^{-1}$ | $n$ $(cm^{-3})$ |
|---|---|---|---|---|---|---|---|---|---|---|
| Si | 0.34 | 289 | $2.9 \times 10^{-2}$ | 1696 | 0.27 | $3.2 \times 10^6$ | 0.84 | 30.3 | 8.37 | $1.98 \times 10^{22}$ |
| n-Si | 0.48 | 352 | $3.2 \times 10^{-2}$ | 1970 | 0.27 | $3.55 \times 10^6$ | 0.75 | 29.6 | 9.07 | $2.52 \times 10^{22}$ |
| n-Si:0.45N | 0.75 | 346 | $3.0 \times 10^{-2}$ | 2065 | 0.27 | $3.3 \times 10^6$ | 0.82 | 32.4 | 9.23 | $2.65 \times 10^{22}$ |
| Si:0.48O | 0.65 | 351 | - | - | - | - | - | - | - | - |

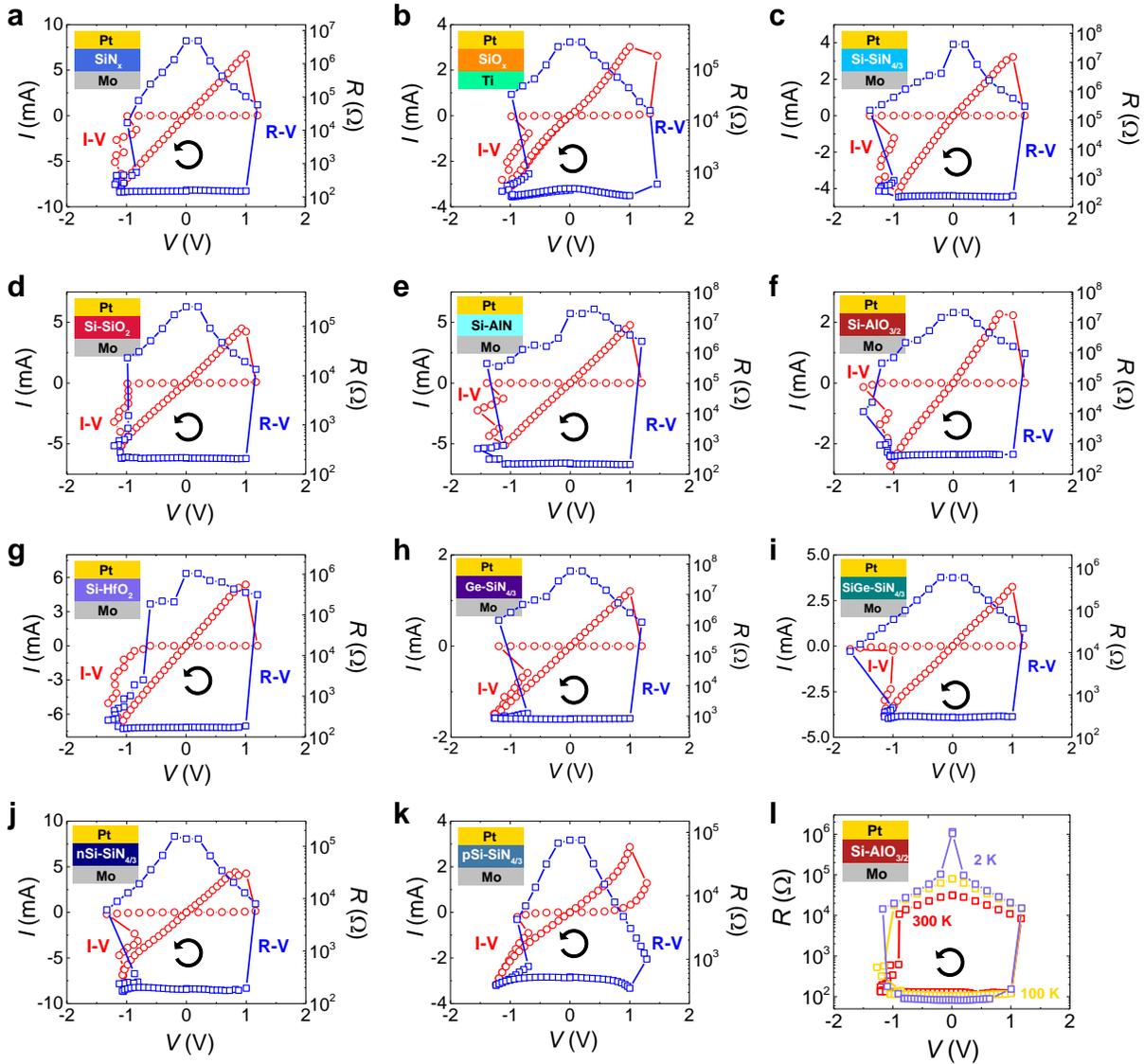

**Extended Data Figure 5 | Similar resistance-voltage (*R-V*) and current-voltage (*I-V*) curves of the reversible metal-insulator transitions around ±1V in various O/N-doped Si/Ge films.** **a,** Si+N; sputtered in $N_2$. **b,** Si+O; sputtered in $O_2$. **c,** Si+N; Si cosputtered with $Si_3N_4$. **d,** Si+O; Si cosputterd with $SiO_2$. **e,** Si+N; Si cosputtered with AlN. **f,** Si+O; Si cosputtered with $Al_2O_3$. **g,** Si+O; Si cosputtered with $HfO_2$. **h,** Ge+N, Ge cosputtered with $Si_3N_4$. **i,** $Si_{0.5}Ge_{0.5}$+N film; $Si_{0.5}Ge_{0.5}$ cosputtered with $Si_3N_4$. **j,** n-Si+N; n-Si cosputtered with $Si_3N_4$. **k,** p-Si+N; p-Si cosputtered with $Si_3N_4$. Above curves were all measured at room temperature. **l,** Transitions of Si+O film (Si cosputtered with $Al_2O_3$) at different temperatures: 300K, 100K and 2K. Arrows indicates transition direction, which is counter-clockwise in all cases. In **a-l**, *V* represents the actual voltage drop on Si films during switching, while the voltage applied to the top electrode is larger (especially in positive bias, with a positive voltage applied to the top Pt electrode) due to the load resistance ~ 500 Ω. See methods for details.

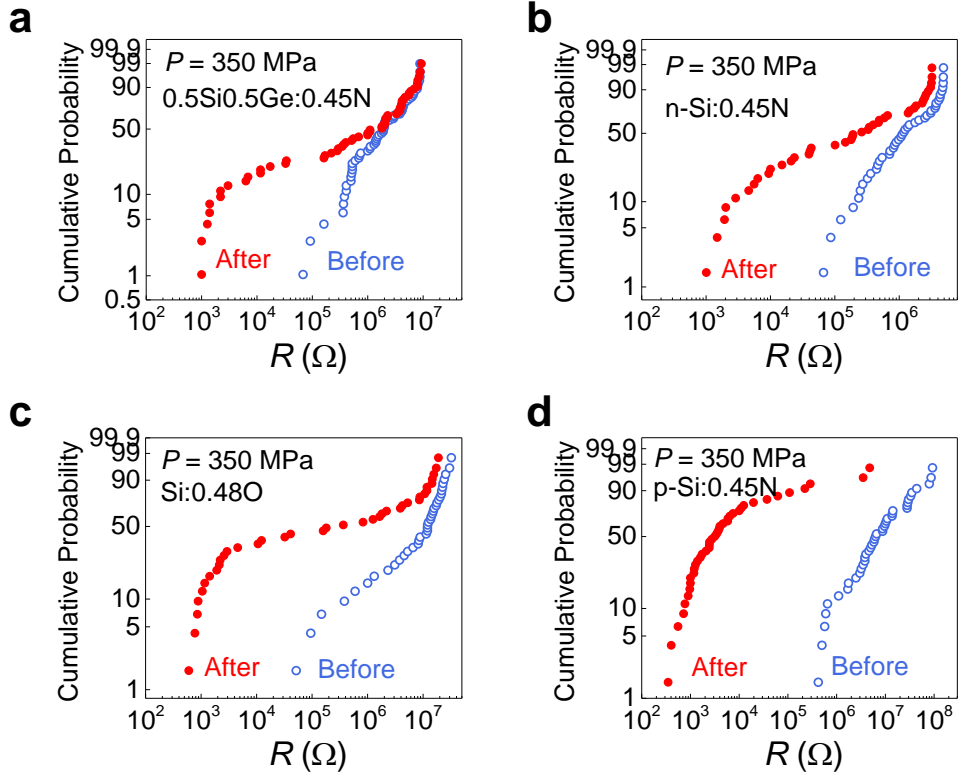

**Extended Data Figure 6 | Insulator-to-metal transition induced by external hydraulic pressure of 350 MPa in various O/N-doped Si films: a,** 0.5Si:0.5Ge:0.45N. **b,** n-Si:0.45N. **c,** Si:0.48O. **d,** p-Si:0.45N. At this pressure and 50% transition yield, the relative resistance drop increases in the order of **a**, **b**, **c**, **d**.

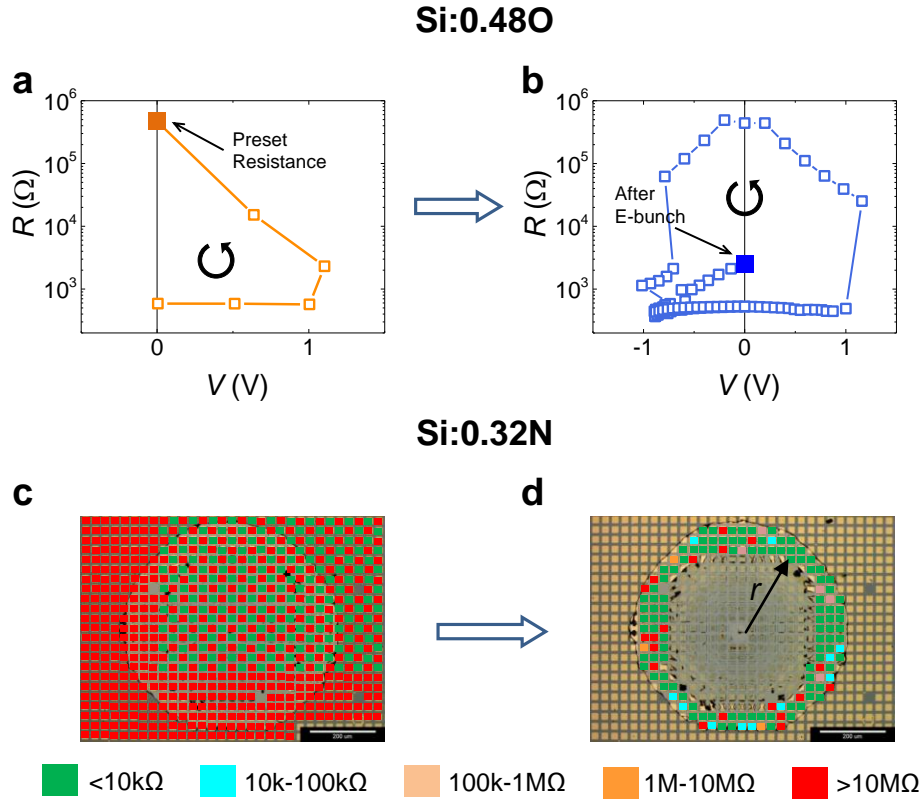

**Extended Data Figure 7 | Insulator-to-metal transition in O/N-doped Si films induced by magnetic pressure delivered by a SLAC 20 GeV electron bunch. a,** Transition curve that pre-set Si:0.48O to insulator state $> 10^5$ $\Omega$, an insulator state, before exposure to electron bunch. **b,** Transition curve of the same cell in **a** after one shot of an electron bunch, which lowered the resistance to $< 10^4$ $\Omega$, a metallic state. In **a-b**, $V$ represents the actual voltage drop on Si films during switching, while the voltage applied to the top electrode is larger (especially in positive bias) due to the load resistance ~ 500 $\Omega$. See methods for details. **c,** Resistance map of Si:0.32N cells (shown as squares) preset to >10 M$\Omega$ (red dot) or <10 k$\Omega$ (green dot) before exposure to electron bunch. The sample was next coated by an insulating photoresist that does not affect magnetic field. **d,** Resistance map of same region as **c** after one shot of an electron bunch at $r$=0. While metallic cells (green) maintained their low resistance, insulator cells (red) were transitioned to either metallic state (green) or intermediate states (orange, blue; see color spectrum at bottom.) Resistance was not read in the central region (blurry grey, $r$<240 µm) where top electrodes were completely torn off by magnetic pressure. Resistance was also not read in the outer region, $r$>300 µm, where photoresist survived and covered the cells. Resistance read by 0.1 V DC. Electron bunch has a size of 20 µm by 20 µm by 20 µm. Lower bound magnetic pressure $P_B$~1,680 MPa at $r$=20 µm at the edge of electron bunch and falls roughly with ~ $1/r^2$, which yields ~48 MPa at $r$=240 µm (edge of blurry zone) and 32 MPa at $r$=300 µm (outer radius of transition region.)

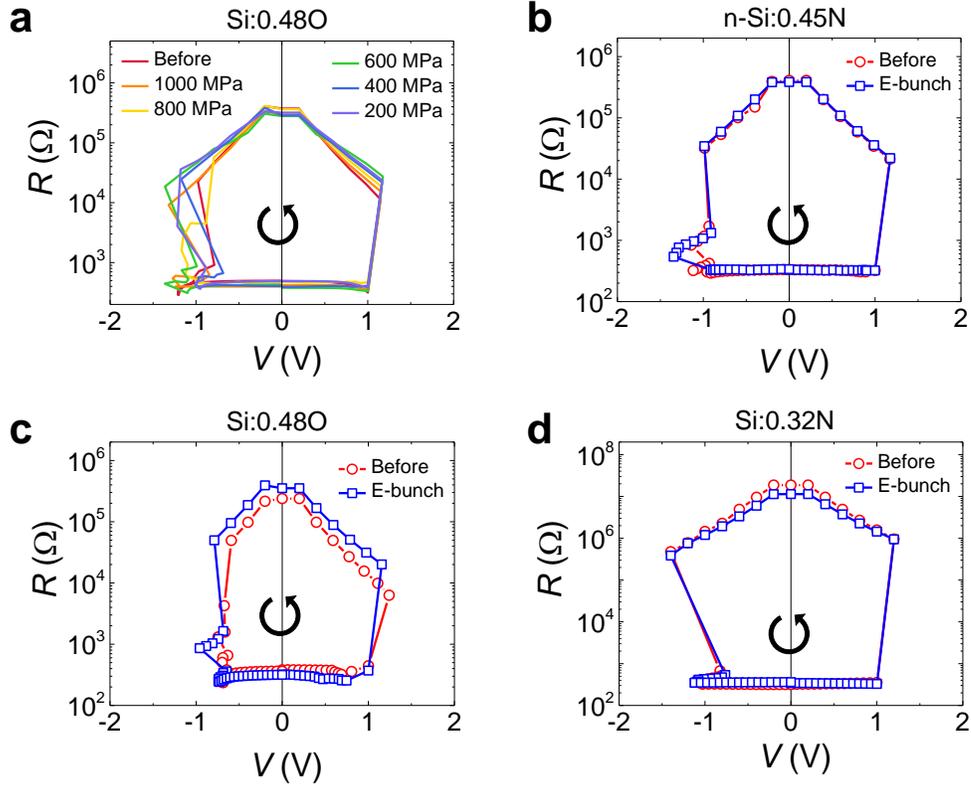

**Extended Data Figure 8 | Comparison of electrical-triggered transitions before and after hydraulic- and magnetic-pressure-triggered insulator-metal transitions in O/N doped Si films. a,** Resistance-voltage ($R$-$V$) curves of Si:0.48O film before and after treatment by 200 MPa-1,000 MPa hydraulic pressure. All the transition curves overlap, indicating no damage to the cells was caused by the pressure treatment. **b,** Resistance-voltage ($R$-$V$) curves of n-Si:0.45N film before and after pressure treatment of 350 MPa. The curves are almost identical. **c,** Resistance-voltage ($R$-$V$) curves of Si:0.48O film before and after exposure to one shot of an electron bunch. The curves are very similar. **d,** Resistance-voltage ($R$-$V$) curves of n-Si:0.32N film before and after exposure to one shot of an electron bunch. The curves are almost identical. Arrows indicate transition directions. In **a-d**, $V$ represents the actual voltage drop on Si films during switching, while the voltage applied to the top electrode is larger (especially in positive bias) due to the load resistance ~ 500 Ω. See methods for details.